\documentclass[aip,apl,amsmath,amssymb,reprint,floatfix]{revtex4-1}
\usepackage{graphicx}
\usepackage{dcolumn}
\usepackage{bm}
\usepackage[utf8]{inputenc}
\usepackage[T1]{fontenc}
\usepackage{mathptmx}
\usepackage{etoolbox}
\usepackage{comment}
\usepackage{xurl}       % Allows line breaks in long URLs/DOIs
\usepackage{hyperref}
\usepackage[markup=highlight,commandnameprefix=always]{changes}
\definecolor{changes@highlight}{RGB}{255,255,0}

\makeatletter
\def\@email#1#2{%
 \endgroup
 \patchcmd{\titleblock@produce}
  {\frontmatter@RRAPformat}
  {\frontmatter@RRAPformat{\produce@RRAP{*#1\href{mailto:#2}{#2}}}\frontmatter@RRAPformat}
  {}{}
}%
\makeatother
\begin{document}

\title{Separating Intrinsic and Domain-Mediated Anomalous Hall Conductivity in Co\texorpdfstring{$_3$Sn$_2$S$_2$}{3Sn2S2} via Contact Engineering}

\author{Eddy Divin Kenvo Songwa}
\author{Shaday Jesus Nobosse Nguemeta}
\author{Hodaya Gabber}
\author{Renana Aharonof}
\author{Dima Cheskis}
\email{dimach@ariel.ac.il}
\affiliation{Physics Department, Ariel University}

\date{\today}

\begin{abstract}
Decoupling the global Berry-curvature contribution to the anomalous Hall conductivity (AHC) from local domain- and texture-related contributions in bulk ferromagnetic Weyl semimetals is difficult in standard measurements. We
address this in a $\sim 670\,\mu\mathrm{m}$-thick Co$_3$Sn$_2$S$_2$ single crystal using a contact architecture that promotes depth-distributed current flow. We find that the AHC depends on the field-enforced domain state: above $\sim 0.3$~T, a single- or few-domain configuration reveals a momentum-space
intrinsic Berry-curvature response, with a crossover near $\sim 125$~K driven by rapid magnetization decrease and reduced magnetic anisotropy. In low-field zero-field-cooled (ZFC) multidomain states, the Hall response is modified by
domain physics, with possible real-space Berry curvature and moderate extrinsic contributions. These results demonstrate contact engineering as a practical, non-invasive strategy for separating the momentum-space intrinsic AHC from
domain-mediated and extrinsic contributions in thick Weyl semimetal crystals.
\end{abstract}

\maketitle
The persistence of topological transport signatures in Weyl semimetals at elevated temperatures has attracted growing interest in recent years. Beyond fundamental research, Weyl semimetals are promising for applications in memory, logic, and sensing devices~\cite{Birch2025}, bulk gating~\cite{Matsuoka2025}, spintronics~\cite{He2024}, and thermoelectric sensors~\cite{CaoAdvMater2026}, all of which require a stable intrinsic anomalous Hall signal above 77~K that is well separated from extrinsic contributions. Below the Curie temperature, itinerant ferromagnetism breaks time-reversal symmetry (TRSB)~\cite{Armitage2018}, while spin--orbit coupling (SOC) generates topologically protected Weyl points~\cite{Yan2017, Burkov2018} that act as Berry-curvature monopoles in momentum space~\cite{Armitage2018, Xu2015, Xiao2010}. The intrinsic AHC is governed by
$\Omega_n^{z}(\mathbf{k})$, weighted by the
Fermi--Dirac distribution~\cite{Xiao2010, Nagaosa2010} (see Supplementary Material, Sec.~S1), and, in the ideal Weyl-metal limit, the AHC $\sigma_{xy}$ is largely controlled by the momentum-space separation between nodes of opposite chirality~\cite{Burkov2014}:

\begin{equation}
\sigma_{xy} = \frac{e^{2}}{h} \sum_{i} \chi_{i},\Delta k_{i}.
\end{equation}

Here, $e$ is the electron charge, $h$ is Planck's constant, $\chi_{i}=\pm1$ is the chirality of the $i$th Weyl node pair, and $\Delta k_{i}$ is the corresponding node separation in momentum space. Both Weyl-node and SOC-gapped nodal-line avoided-crossing mechanisms~\cite{Nagaosa2010, Yao2004,Liu2022} contribute to the giant AHE in Co$_3$Sn$_2$S$_2$~\cite{He2025, PateAdvSci2024, Rossi2021, Lv2025, Wang2025, PatePRB2023, Zivkovic2022, Orlova2025, Wang2024}, with small interband gaps strongly enhancing the Berry curvature. The dominant intrinsic source remains debated~\cite{Wang2018, Schilberth2023PRB, Liu2018, Minami2020}, and because both the Weyl-node separation $\Delta k$ and the avoided-crossing Berry curvature depend on magnetization~\cite{Burkov2014, He2025, Minami2020}, their finite-temperature contributions are difficult to disentangle~\cite{Rossi2021}.

Beyond intrinsic effects, extrinsic mechanisms such as skew scattering~\cite{Yang2019DomainWallSkew} and side-jump~\cite{Berger1972SideJump} may also contribute to the AHE~\cite{Nagaosa2010}, scaling as $\sigma_{xx}$ and $\sigma_{xx}^2$, respectively. Although chemical doping can tune the Fermi level, it further entangles intrinsic and extrinsic contributions~\cite{He2024ClDopedCo3Sn2S2, Shen2020FeDopedCo3Sn2S2, Onoda2006, Tian2009, Shen2022IntrinsicAHECo3Sn2S2}. Instead, we employ a contact-engineering approach that forms deep ohmic contacts without altering the bulk electronic structure, enabling a cleaner separation of momentum-space intrinsic contributions from other possible intrinsic and extrinsic contributions.

We focus on Co$_3$Sn$_2$S$_2$, a prototypical magnetic Weyl semimetal exhibiting giant AHC and serving as an ideal platform for topological transport studies~\cite{Liu2018,Wang2018,Zhu2023}. Its magnetic properties, however, remain debated. Co atoms form kagome triangles with competing exchange interactions and intrinsic geometric frustration. Along the easy $c$-axis, strong SOC-induced anisotropy stabilizes a collinear itinerant ferromagnetic (FM) state that defines the dominant Berry-curvature chirality, whereas the $ab$-plane magnetism is considerably more complex and remains unresolved.

In particular, several authors have proposed an in-plane antiferromagnetic (AFM) phase as a local rather than globally ordered state. Guguchia \textit{et al.}~\cite{Guguchia2020} reported a reduction of the ferromagnetic volume fraction above $\sim$90~K, suggesting coexistence between a ferromagnetic phase and a competing magnetic state, interpreted as a $120^{\circ}$ in-plane AFM component but without direct evidence for long-range order. Consistently, neutron diffraction measurements~\cite{Soh2022} found no evidence of long-range in-plane AFM order, a result that is inherently difficult to obtain by standard diffraction techniques because the relevant magnetic reflections coincide with nuclear Bragg peaks~\cite{Ekahana2025}. At the same time, nuclear magnetic resonance (NMR) measurements~\cite{Mukhamedshin2025} reveal a continuous evolution of the local magnetic field with no distinct spectral signatures of phase separation, arguing against clear FM--AFM two-phase coexistence, although they remain consistent with weak or short-range AFM-related effects, including spin canting, noncoplanar spin textures, and domain-wall magnetism.

Exchange-biased Hall loops and ``bow-tie'' hysteresis observed under field-cooled (FC) conditions have been attributed to frustration-induced coexistence of ferromagnetic and glassy phases~\cite{Lachman2020}, whereas alternative interpretations show that similar features can arise within a single ferromagnetic phase through a multidomain intermediate during magnetization reversal and a secondary spin population governing magnetic memory effects~\cite{Menil2025}. Consistently, Kerr-microscopy measurements reveal that the anomaly is localized 
within domain walls~\cite{Lee2022}, while transport studies attribute the feature near the anomaly temperature $T_A \sim 130$~K to anomalous depinning of magnetic domain walls within the ferromagnetic phase, rather than to a conventional 
thermodynamic phase transition~\cite{Shen2023}. Additional support comes from angle-dependent magnetization measurements, which are incompatible with a spin-glass scenario~\cite{Zivkovic2022}. Within this picture, the magnetic state under ZFC is naturally described as a multidomain configuration in which `up'' and `down'' $c$-axis domains coexist. Domain walls separating these domains host locally modified, often noncollinear spin textures that can introduce additional Berry-curvature-related and scattering contributions absent in the single-domain state~\cite{Lee2022, Schilberth2025, Nagaosa2010}. Such textures may generate finite spin chirality and associated real-space Berry-phase effects~\cite{Nagaosa2010}. In this regime, Pate~\textit{et al.}~\cite{PateAdvSci2024, PatePRB2023, PatePhD2024} demonstrated that domain dynamics lead to ``bow-tie'' hysteresis and strong suppression of the anomalous Hall conductivity due to cancellation between oppositely magnetized domains, while the field required to restore a single-domain state exhibits a pronounced maximum near $T_A \sim 130$~K.

From the transport perspective, the anomalous Hall conductivity reflects both intrinsic and extrinsic mechanisms. The intrinsic contribution is governed by momentum-space Berry curvature from Weyl nodes and
SOC-gapped nodal-line avoided crossings. Additional contributions may arise from noncollinear or noncoplanar spin textures, which generate real-space Berry curvature and a topological Hall effect via an effective magnetic field acting on conduction electrons~\cite{Neubauer2009}. Domain
walls may further contribute through skew scattering, side-jump, and effects related to chiral gauge fields and SOC-gap
modulation~\cite{SornPRB2021, OzawaJPSJ2024, Kobayashi2022, CaoAdvMater2026}.
Importantly, applying a sufficiently large magnetic field ($\gtrsim 0.2$--$0.5$~T), consistent with the field scale at which multidomain states collapse in Co$_3$Sn$_2$S$_2$~\cite{PateAdvSci2024, PatePRB2023,Noah2022}, suppresses the multidomain configuration and stabilizes a nearly single-domain state. In this regime, domain-related extrinsic contributions are strongly reduced, and the measured AHC approaches the intrinsic value governed by the Berry curvature of the Weyl band structure.

Motivated by these prior results, we performed experiments on Co$_3$Sn$_2$S$_2$ aimed at enhancing local effects to better elucidate the nature of the AHC, particularly in the $90$--$150~\mathrm{K}$ regime where multiple competing processes have been reported. Measurements were carried out on a $670~\mu\mathrm{m}$-thick single crystal, comparable to bulk samples reported in the literature. Structural and compositional characterization (SEM, EDS, XRD) confirmed high crystal quality and a well-defined crystallographic orientation along the $c$-axis (see Supplementary Material, Sec.~S2).

\begin{figure}[htbp]
\centering
\includegraphics[width=1\linewidth]{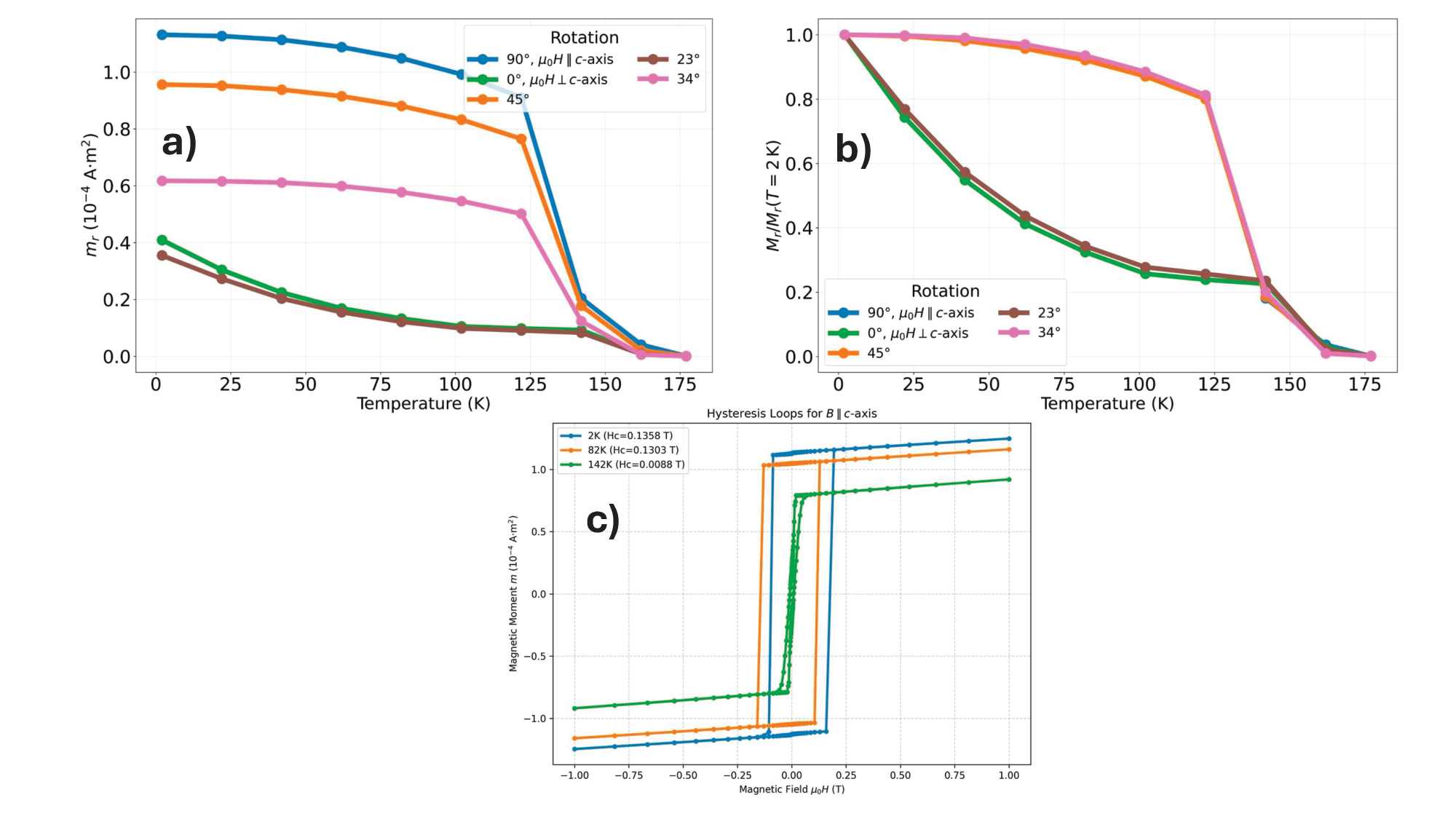}
\caption{Magnetization for different orientations of the applied field:
(a) magnetic moment (proportional to absolute magnetization);
(b) magnetization normalized to its value at $T = 2\,\mathrm{K}$.
$\theta = 0^\circ$ corresponds to $\mathbf{\mu_0 H} \perp \mathbf{c}$ and $\theta = 90^\circ$ to $\mathbf{\mu_0 H} \parallel \mathbf{c}$.}
\label{fig:magnetization}
\end{figure}

Prior to transport measurements, magnetization was characterized to establish the magnetic anisotropy. The data reveal a pronounced easy axis along the $c$-axis (Fig.~\ref{fig:magnetization}). In our angular convention, $\theta = 0^\circ$ corresponds to the applied field directed perpendicular to the $c$-axis (i.e., parallel to the sample surface), following the geometry of the Magnetic Property Measurement System (MPMS-3) magnetometer.

In Fig.~\ref{fig:magnetization}, it is shown that the magnetic field is rotated away from the $c$-axis, the absolute magnetization decreases while the normalized curves remain nearly unchanged, consistent with strong out-of-plane anisotropy in Co$_3$Sn$_2$S$_2$. 

\begin{figure}[htbp]
\centering
\includegraphics[width=0.30\linewidth]{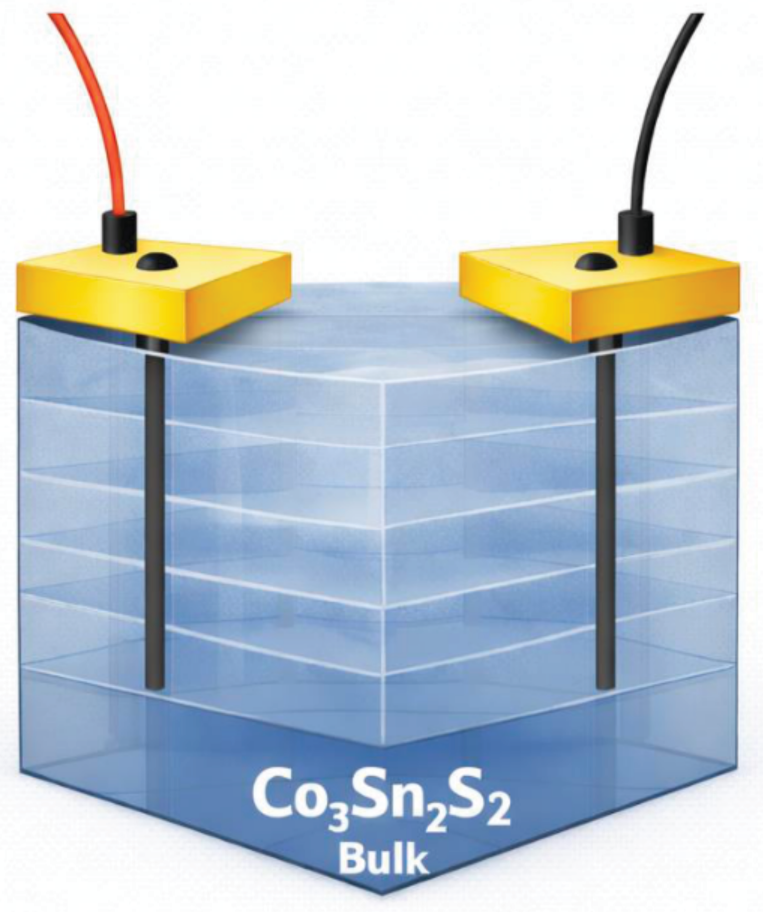}
\caption{Electrical contact geometry: electrical contacts formed by FIB drilling and tungsten (W) filling, connecting the gold pads to the crystal.}
\label{fig:contacts}
\end{figure}

To promote current excitation across multiple depth layers, low-resistance ohmic contacts were engineered by FIB-assisted deposition of tungsten-filled contacts through gold surface pads into the crystal bulk (Fig.~\ref{fig:contacts}). Their resistance is significantly lower than that of the bulk crystal, ensuring depth-distributed current injection; fabrication details and resistance estimates are provided in the Supplementary Material (Sec.\ S3).

The anomalous Hall conductivity $\sigma_{xy}$ is obtained by tensor inversion of the measured resistivities, with the ordinary Hall contribution $R_0 B$ subtracted using the high-field linear slope of $-\rho_{xy}$; the derivation of $\rho_{xx}$ and its validity under moderate conductivity anisotropy are detailed in the Supplementary Material (Secs. S4 and S5).

\begin{figure}[htbp]
\centering
\includegraphics[width=\linewidth]{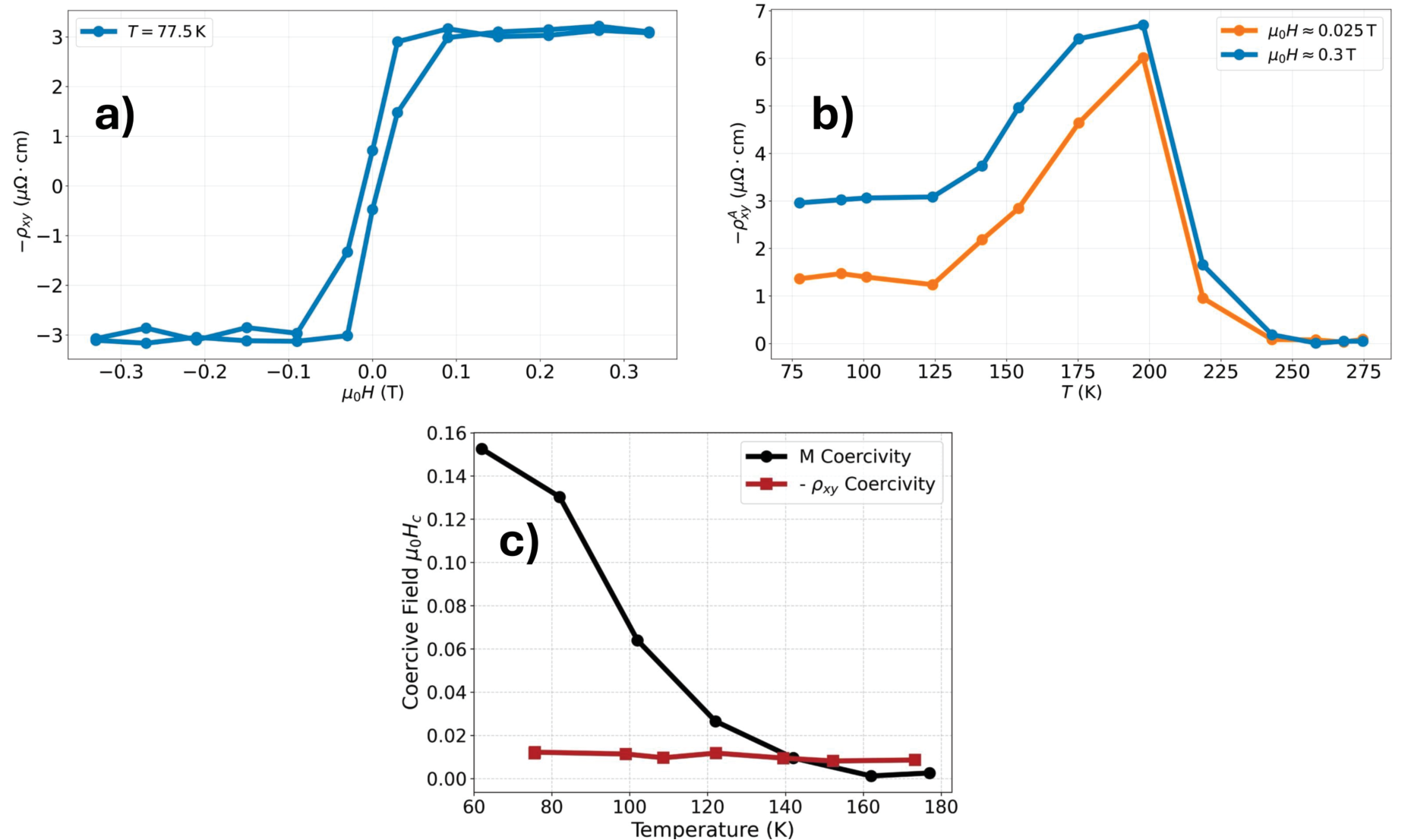}
\caption{(a) Magnetic hysteresis loop at $T = 77\,\mathrm{K}$;
(b) corresponding Hall resistivity $-\rho_{xy}$ measured at two applied fields, $\mu_0 H = 0.025\,\mathrm{T}$ (orange) and $\mu_0 H = 0.3\,\mathrm{T}$ (blue);
(c) coercive fields extracted from the magnetization and $-\rho_{xy}$ loops.
$\mu_0 \mathbf{H} \parallel \mathbf{c}$.}
\label{fig:hysteresis}
\end{figure}

Hall measurements were performed using a Lake Shore Cryotronics M91 FastHall measurement system with an external electromagnet for hysteresis-loop acquisition. All measurements were done with $\mathbf{\mu_0 H} \parallel \mathbf{c}$-axis. The sample was cooled to 77~K with liquid nitrogen and slowly warmed to room temperature under vacuum, allowing complete hysteresis-loop recording at successive temperatures. Figure~\ref{fig:hysteresis}a) shows a representative $-\rho_{xy}$ hysteresis loop, and Fig.~\ref{fig:hysteresis}b) presents its temperature dependence at $\mu_0 H = 0.025,\mathrm{T}$ and at $\mu_0 H = 0.3,\mathrm{T}$ when The ordinary Hall contribution $R_0 B$ remains below 1\%. 

\begin{figure}[h]
    \centering
    \includegraphics[width=\linewidth]{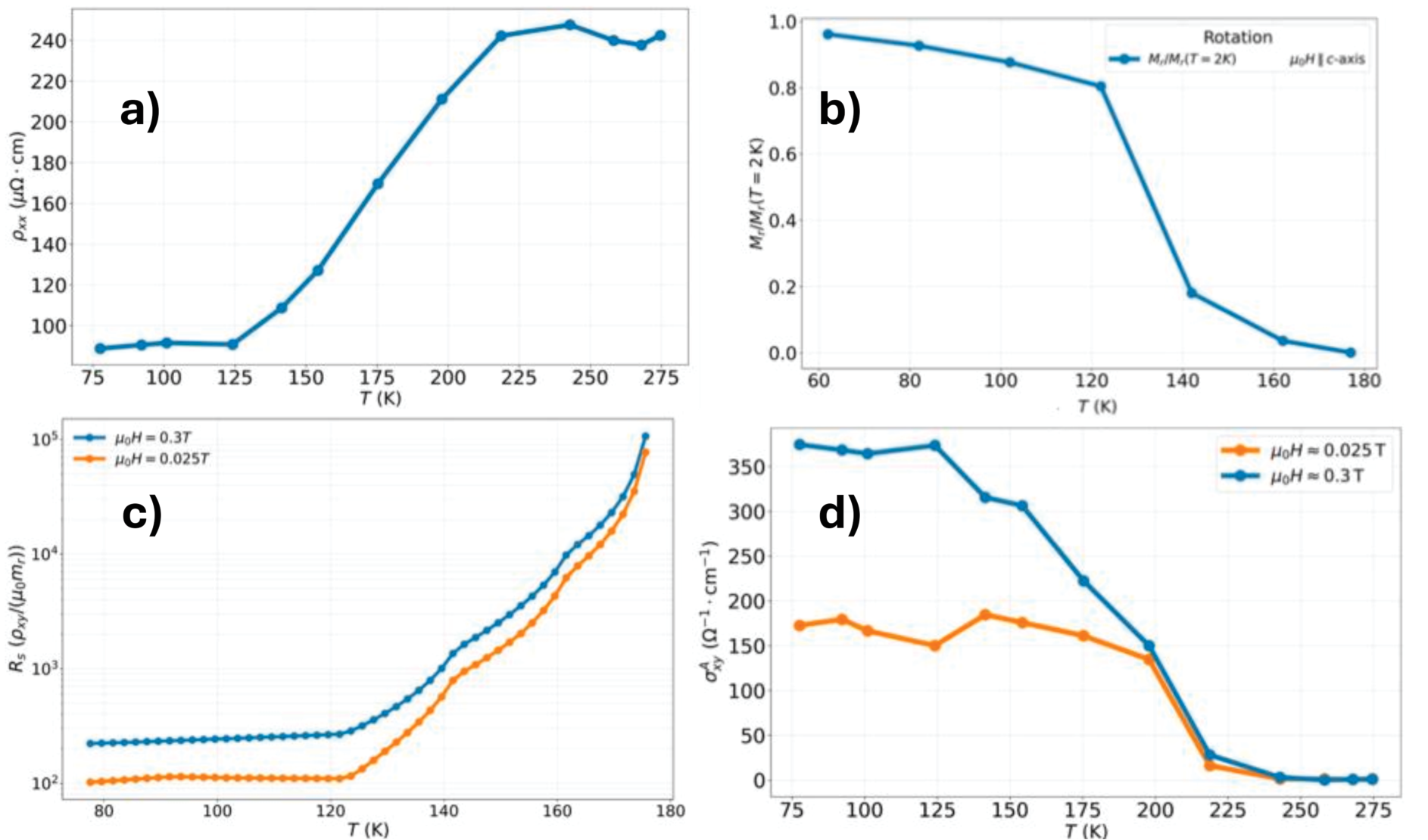}
    \caption{Transport measurements:
(a) longitudinal resistivity $\rho_{xx}$;
(b) magnetization normalized to its value at $T = 2\,\mathrm{K}$;
(c) interpolated Hall coefficient $R_s$;
(d) anomalous Hall conductivity $\sigma_{xy}$ measured at
$\mu_0 H = 0.025\,\mathrm{T}$ (orange) and
$\mu_0 H = 0.3\,\mathrm{T}$ (blue).
$\mu_0 \mathbf{H} \parallel \mathbf{c}$.}
    \label{fig:transport}
\end{figure}

The $-\rho_{xy}$ reversal is substantially smoother than the magnetization loop. Comparing the coercive fields extracted from both (Fig.~\ref{fig:hysteresis}c) reveals a pronounced discrepancy: the magnetization coercivity decreases
significantly above 70~K, while the Hall coercivity remains small and nearly temperature independent. The longitudinal resistivity begins to increase near 125~K (Fig.~\ref{fig:transport}a), simultaneously with a decrease in
magnetization above $\sim$120~K (Fig.~\ref{fig:transport}b). The anomalous Hall coefficient $R_s$, interpolated from $-\rho_{xy}$ and $M$ (Fig.~\ref{fig:transport}c), shows an upturn near 120~K. The resulting $\sigma_{xy}$ (Fig.~\ref{fig:transport}d) at $\mu_0 H = 0.3\,\mathrm{T}$
tracks the magnetization closely, whereas at $\mu_0 H = 0.025\,\mathrm{T}$ it remains nearly constant up to ${\sim}\,150\,\mathrm{K}$ before decreasing slowly.

\begin{figure}[h]
    \centering
    \includegraphics[width=1\linewidth]{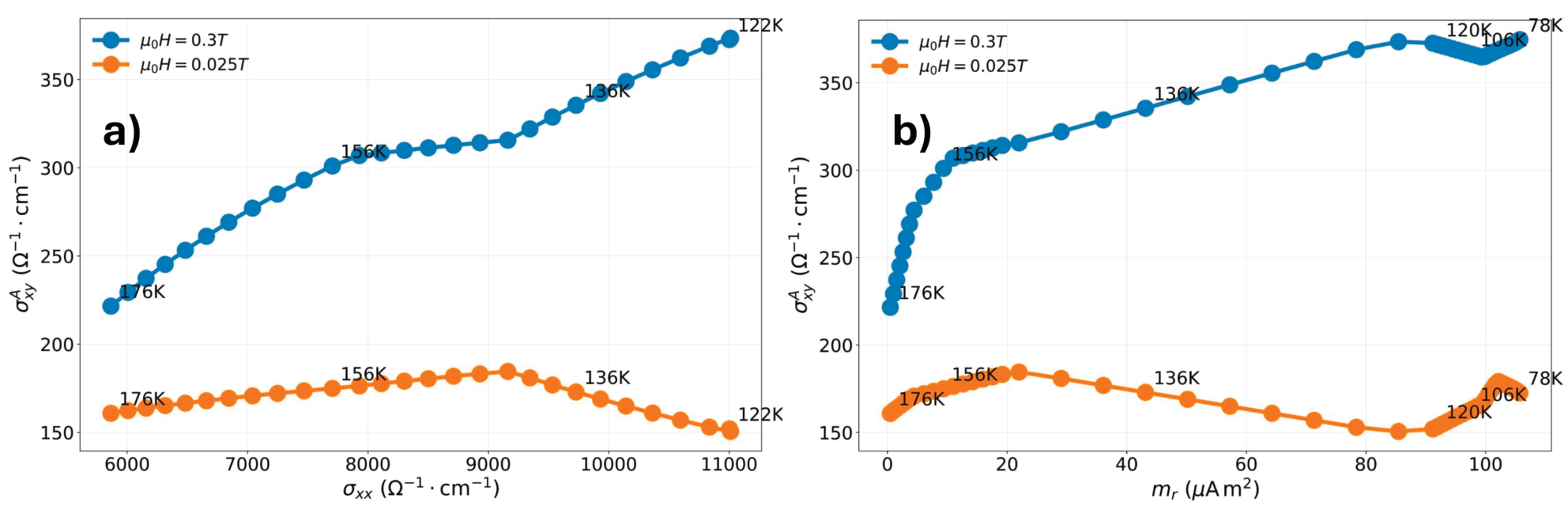}
    \caption{
(a) Anomalous Hall conductivity $\sigma_{xy}$ as a function of longitudinal
conductivity $\sigma_{xx}$, illustrating scaling behavior and enabling
identification of intrinsic and extrinsic contributions;
(b) anomalous Hall conductivity $\sigma_{xy}$ as a function of magnetic
moment $m_r$, demonstrating the correlation between Berry-curvature-driven
transport and magnetic order.
$\mu_0 \mathbf{H} \parallel \mathbf{c}$.}
    \label{fig:sigmaxyassigmaxx}
\end{figure}

Scaling analysis of $\sigma_{xy}$ versus $\sigma_{xx}$
(Fig.~\ref{fig:sigmaxyassigmaxx}a) and versus $M_r$
(Fig.~\ref{fig:sigmaxyassigmaxx}b) was performed following
Refs.~\cite{Nagaosa2010, Onoda2006, Tian2009}. At $\mu_0 H = 0.3\,\mathrm{T}$,
$\sigma_{xy}$ decreases nearly linearly with $\sigma_{xx}$ above 120~K, and
$\sigma_{xy}(M_r)$ decreases linearly as well. At $\mu_0 H = 0.025\,\mathrm{T}$, the behavior is more complex: despite decreasing magnetization, $\sigma_{xy}$ initially increases slightly up to $\sim$140~K, with a non-monotonic
$\sigma_{xy}(\sigma_{xx})$ dependence, and $\sigma_{xy}(M_r)$ increases above 120~K even as the magnetization is strongly reduced. Above $\sim$140~K, $\sigma_{xy}$ again decreases linearly with $\sigma_{xx}$.

Below $\sim$125~K, the AHC at $\mu_0 H = 0.3$~T is predominantly intrinsic, with high, nearly constant $\sigma_{xy}$ and weak scattering dependence, consistent with a single- or few-domain state aligned along the $c$-axis. Above $\sim$125~K, however, $\sigma_{xy}$ decreases with magnetization, and its partial scaling with $\sigma_{xx}$ indicates an additional skew-scattering contribution.

At $\mu_0 H = 0.025$~T, $\sigma_{xy}$ remains nearly constant between 120 and 140~K despite the decrease in longitudinal conductivity, the reduction in magnetization, and the weakening of magnetic anisotropy. We attribute this behavior to a multidomain state in which a possible real-space Berry-curvature contribution, associated with canted in-plane AFM order and non-coplanar spin textures within domain walls, helps sustain the Hall response. This interpretation is supported by the angle-dependent magnetization measurements. For angles below $56^\circ$ from the $c$-axis ($34^\circ$ from the sample surface), a pronounced temperature-dependent reduction in magnetization is observed across all measured field directions. In contrast, for angles above $67^\circ$ from the $c$-axis 
($23^\circ$ from the sample surface), the magnetization curves change only weakly up to 150~K, exhibiting at most a small hump before converging onto a common trace for all directions near 177~K. Enhanced extrinsic contributions from skew scattering and 
side-jump processes are also expected in this regime.

In conclusion, the key distinction between the high- and low-field regimes lies in the domain state enforced by the applied field. Only above $\sim$0.3~T, consistent with the observations of Noah~\textit{et~al.} and Pate~\textit{et~al.}, does the system reach a single- or few-domain state. This regime reveals the momentum-space intrinsic Berry-curvature response and enables a clear separation from the much smaller real-space intrinsic and extrinsic contributions to the AHC. In low fields under ZFC conditions, the Hall conductivity is supported by real-space Berry curvature and a moderate extrinsic component. The onset of rapid magnetization decrease and the loss of strong magnetic anisotropy near 125~K drive the reduction in intrinsic AHC observed in the high-field regime, in agreement with calculations of reduced anisotropy in this temperature range. A natural next step is to optimize the contact geometry to further enhance the intrinsic AHC signal while preserving this field-regime separation.

\begin{acknowledgments}
The authors thank Professor Yuri Gorodetski for fruitful discussions and for fabricating the specialized FIB-assisted contacts, Dr.\ Victor Shelukhin of the Weizmann Institute of Science for assistance in preparing the test setup for the Lake Shore Cryotronics FastHall measurement system, and Dr.\ Anna Eyal of the Quantum Materials Center, Technion--Israel Institute of Technology, for assistance with the magnetization measurements performed using the Quantum Design MPMS-3 system. This work was supported by the Israeli Ministry of Energy under Agreement No.~222-11-010 and Grant No.~318844.
\end{acknowledgments}

\sloppy
\bibliographystyle{aipnum4-1}
\bibliography{topol_transport_resubmission}

\end{document}

\title{Supplementary Material: Separating Intrinsic and Domain-Mediated Anomalous Hall Conductivity in Co\texorpdfstring{$_3$Sn$_2$S$_2$}{3Sn2S2} via Contact Engineering}

\author{Eddy Divin Kenvo Songwa}
\author{Shaday Jesus Nobosse Nguemeta}
\author{Hodaya Gabber}
\author{Renana Aharonof}
\author{Dima Cheskis}
\email{dimach@ariel.ac.il}
\affiliation{Physics Department, Ariel University}

\date{\today}

\maketitle

\section*{S1. Kubo-Formula Expression for the Intrinsic Anomalous Hall
Conductivity}

The intrinsic AHC is governed by the Berry curvature $\Omega_n^{z}(\mathbf{k})$
weighted by the Fermi--Dirac distribution $f_n$~\cite{Xiao2010, Nagaosa2010}:
\begin{equation}
  \sigma_{xy}^{\mathrm{AHE}}(T) = -\frac{e^2}{\hbar} \sum_n \int_{\mathrm{BZ}}
  \frac{d^3\mathbf{k}}{(2\pi)^3} \, f_n(\mathbf{k},T)\, \Omega_n^{z}(\mathbf{k}),
  \label{eq:sigmaxy_intrinsic}
\end{equation}
where the general expression for the Berry curvature is:
\begin{equation}
\Omega_{n}^{z}(\mathbf{k}) = -2\sum_{m \neq n}
\frac{\mathrm{Im}\left[
\langle u_{n\mathbf{k}} \vert \hat{v}_x \vert u_{m\mathbf{k}} \rangle
\langle u_{m\mathbf{k}} \vert \hat{v}_y \vert u_{n\mathbf{k}} \rangle
\right]}{(E_n - E_m)^2}.
\end{equation}
Here $\Omega_{n}^{z}(\mathbf{k})$ is the $z$-component of the Berry curvature
of band $n$, the sum $\sum_{m \neq n}$ runs over all interband contributions,
$\mathrm{Im}[\ldots]$ denotes the imaginary part of the velocity matrix elements
$\hat{v}_{x,y}$, and the $(E_n - E_m)^2$ denominator causes strong enhancement
near avoided crossings such as Weyl points and small SOC-induced gaps.

Near each Weyl node, the occupied band carries a Berry-curvature monopole:
\begin{equation}
  \boldsymbol{\Omega}(\mathbf{q}) = \chi\,\frac{\mathbf{q}}{2|\mathbf{q}|^{3}},
\end{equation}
where $\mathbf{q}=\mathbf{k}-\mathbf{k}_W$ and $\chi=\pm1$ is the chirality.
Integrating the Chern number $C(k_z)$ of successive BZ slices --- which jumps
by $\chi_i$ at each node --- yields~\cite{Burkov2014}:
\begin{equation}
  \sigma_{xy} = \frac{e^{2}}{h}\sum_{i}\chi_{i}\,\Delta k_{i},
\end{equation}
so the intrinsic AHC is directly proportional to the momentum-space separation
between nodes of opposite chirality. At finite temperature, this intrinsic
contribution is further modulated by the Fermi--Dirac factor
$f_n(\mathbf{k},T)$ in Eq.~(\ref{eq:sigmaxy_intrinsic}), which smoothly
suppresses the Berry-curvature integral as thermally excited states depopulate
the relevant bands near the Weyl nodes.

\section*{S2. Structural and Compositional Characterization}

Structural and compositional characterization was performed using scanning
electron microscopy (SEM) to assess the surface morphology, energy-dispersive
X-ray spectroscopy (EDS) to confirm the elemental composition, and X-ray
diffraction (XRD) to verify the crystallographic structure and orientation.

\begin{figure}[htbp]
\centering
\includegraphics[width=0.75\linewidth]{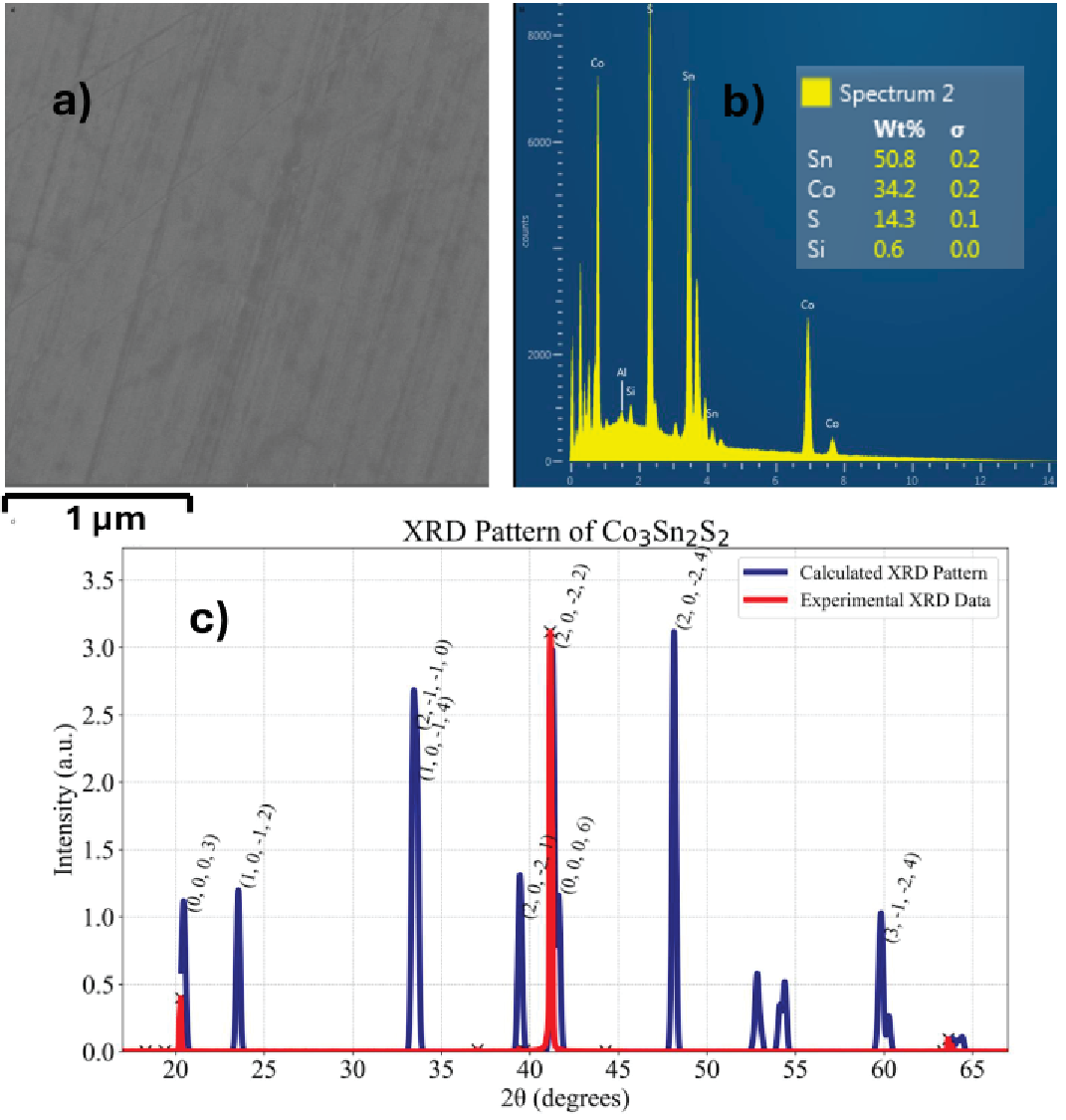}
\caption{Structural and compositional characterization of the
Co$_3$Sn$_2$S$_2$ sample: (a) SEM image, (b) EDS spectrum, and
(c) XRD pattern.}
\label{fig:sem_eds_xrd_supp}
\end{figure}

From these results, we conclude that the crystal is of high quality and that
the $c$-axis is perpendicular to the sample surface, confirming the intended
crystallographic orientation for transport measurements with
$\mu_0\mathbf{H} \parallel \mathbf{c}$.

\section*{S3. Electrical Contact Fabrication and Resistance Estimate}

To enable reliable transport measurements across the full thickness of the
crystal, high-quality electrical contacts were fabricated using a focused ion
beam (FIB) technique. Gold pads (200~nm thick) were first deposited on the
crystal surface by sputtering. The FIB was then used to drill channels through
the gold pads and into the crystal bulk, and the resulting holes were filled
with tungsten (W) by FIB-assisted chemical vapor deposition, forming W
nanowires that connect the surface pads to deeper crystal layers.

The use of FIB-deposited contacts for establishing reliable low-resistance
ohmic connections to crystalline materials is well
established~\cite{Chen2015JOVE, Murata2013NRL, Ke2012APL}.
The typical measured resistivity of FIB-deposited tungsten is
$\sim 200\,\mu\Omega\cdot\mathrm{cm}$~\cite{Sadki2004SuperconductingW,
Niles2008FIBViaResistance, Horvath2006FIBWires}, which is significantly lower
than that of Co$_3$Sn$_2$S$_2$. Our resistance estimates show that an
FIB-filled W contact (50~nm diameter, 16~$\mu$m depth) has a resistance much
lower than that of the corresponding bulk crystal segment, ensuring that the
W-filled channels act as low-impedance current injectors rather than
bottlenecks.

The geometry of the resulting current paths is illustrated in Fig.~2 of the
main text.

\section*{S4. Conversion from Hall Resistivity to Hall Conductivity}

The anomalous Hall conductivity is obtained via the standard tensor
inversion~\cite{Nagaosa2010}:
\begin{equation}
  \sigma_{xy} = -\frac{\rho_{xy}}{\rho_{xx}^2 + \rho_{xy}^2}.
\end{equation}
The longitudinal resistivity is accessed through the sheet resistance
$R_\square$ measured in the van der Pauw configuration:
\begin{equation}
  \rho_{xx} = R_\square\, t,
\end{equation}
where $t \approx 670\,\mu\mathrm{m}$ is the crystal thickness. This relation
is exact for an isotropic conductor. For Co$_3$Sn$_2$S$_2$, the in-plane to
out-of-plane conductivity ratio is $\sigma_{ab}/\sigma_c \approx
10$~\cite{Schilberth2023PRB}, indicating moderate anisotropy; however, as
shown by Tanatar \textit{et al.}~\cite{Tanatar2009PRB} for layered conductors
with comparable anisotropy, $\rho_{xx} = R_\square\, t$ remains approximately
valid for predominantly in-plane current flow.

\section*{S5. Separation of Ordinary and Anomalous Hall Contributions}

The total Hall resistivity contains ordinary (Lorentz) and anomalous
contributions~\cite{Nagaosa2010}:
\begin{equation}
  -\rho_{xy} = R_0 B + R_s \mu_0 M,
\end{equation}
where $R_0$ and $R_s$ are the ordinary and anomalous Hall coefficients, $B$
is the applied flux density, and $M$ is the magnetization. To isolate the
anomalous part, $R_0$ is extracted by linear regression of $-\rho_{xy}$
versus $B$ at high fields where $M$ is saturated and $R_s\mu_0 M$ is
constant. The anomalous Hall resistivity
\begin{equation}
  -\rho_{xy}^{\mathrm{AHE}} = R_s \mu_0 M
\end{equation}
is then obtained by subtracting $R_0 B$ from the total $-\rho_{xy}$ at each
temperature and field. The ordinary contribution remains below $1\%$ at all
fields, confirming that the anomalous component dominates in all cases.

\sloppy
\bibliographystyle{aipnum4-1}
\bibliography{topol_transport_resubmission}